\documentclass[twocolumn,amsmath,amssymb,pra]{revtex4-1}

\usepackage{graphicx}
\usepackage{bm}
\usepackage[usenames,dvipsnames,svgnames]{xcolor}
\usepackage[breaklinks=true,colorlinks=true,linkcolor=blue,urlcolor=blue,citecolor=blue]{hyperref}

\usepackage{dcolumn}
\newcolumntype{d}{D{.}{.}{-1}}

\usepackage{tabularx}
\newcommand{\ab}[0]{\emph{ab initio} }
\newcolumntype{C}[1]{>{\setlength\hsize{#1\textwidth}\centering}X}
\newcolumntype{L}[1]{>{\setlength\hsize{#1\textwidth}\raggedright}X}

\begin{document}

\title{Long-range interactions between polar bialkali ground-state molecules in arbitrary vibrational levels}

\author{R. Vexiau, M. Lepers, M. Aymar, N. Bouloufa-Maafa and O. Dulieu}
\affiliation{Laboratoire Aim\'e Cotton, CNRS/Universit\'e~Paris-Sud/ENS-Cachan, B\^at.~505, Campus d'Orsay, 91405 Orsay, France}
\email{maxence.lepers@u-psud.fr}

\date{\today}

\begin{abstract}
We have calculated the isotropic $C_6$ coefficients characterizing the long-range van der Waals interaction between two identical heteronuclear alkali-metal diatomic molecules in the same arbitrary vibrational level of their ground electronic state $X^1\Sigma^+$. We consider  the ten species made up of $^7$Li, $^{23}$Na, $^{39}$K, $^{87}$Rb and $^{133}$Cs. Following our previous work [M.~Lepers \textit{et.~al.}, Phys.~Rev.~A \textbf{88}, 032709 (2013)] we use the sum-over-state formula inherent to the second-order perturbation theory, composed of the contributions from the transitions within the ground state levels, from the transition between ground-state and excited state levels, and from a crossed term. These calculations involve a combination of experimental and quantum-chemical data for potential energy curves and transition dipole moments. We also investigate the case where the two molecules are in different vibrational levels and we show that the Moelwyn-Hughes approximation is valid provided that it is applied for each of the three contributions to the sum-over-state formula. Our results are particularly relevant in the context of inelastic and reactive collisions between ultracold bialkali molecules, in deeply bound or in Feshbach levels.
\end{abstract}

\maketitle

\section{Introduction}
\label{sec:intro}

In the field of ultracold matter referring to dilute gases with a kinetic energy $E=k_BT$ equivalent to a temperature $T$ well below 1~mK, dipolar atomic and molecular systems are currently attracting a considerable interest, as they offer the possibility to study highly-correlated systems, many-body physics, quantum magnetism, with an exceptional level of control \cite{baranov2008,lahaye2009}. Essentially dipolar gases are composed of particles carrying a permanent electric and/or magnetic dipole moment, which induces anisotropic long-range dipole-dipole interactions between particles, that can be designed at will using external electromagnetic fields. For example dipole-dipole interactions were shown to modify drastically the stereodynamics of bimolecular reactive collisions at ultralow temperature \cite{demiranda2011,quemener2012}.

In this perspective the recent production of ultracold heteronuclear alkali-metal diatomic molecules in the lowest rovibronic \cite{ni2008,deiglmayr2008,aikawa2010,molony2014,shimasaki2015} and even hyperfine level \cite{ospelkaus2010b,takekoshi2014} is very promising. A crucial step common to most experiments is the conversion of ultracold atom pairs into so-called Feshbach weakly-bound molecules \cite{kohler2006}, which are then transferred to a desired target state using a coherent laser-assisted process known as Stimulated Raman Adiabatic Passage (STIRAP) \cite{bergmann1998}. This target state is often the lowest rovibrational or hyperfine one, but in principle it can be any molecular level \cite{ospelkaus2008, deiss2015}. Such investigations have stimulated a wealth of combined theoretical and spectrocopic studies devoted to various polar bialkali molecules \cite{stwalley2004,tscherneck2007,docenko2007,zaharova2009,stwalley2010,debatin2011,klincare2012,schulze2013,borsalino2014,patel2014,borsalino2015}. 

Following Ref.~\cite{zuchowski2010} heteronuclear bialkali molecules AB are usually identified as {}``reactive" or {}``non-reactive", depending on whether the reaction
\begin{equation}
  2\mathrm{AB}\to\mathrm{A}_2+\mathrm{B}_2
\label{eq:chem-reac}
\end{equation}
is exothermic or endothermic. This classification, which is based on the differences in dissociation energies at the equilibrium distance of $\mathrm{AB}$, $\mathrm{A}_2$ and $\mathrm{B}_2$, is not modified by the effect of zero-point energy for molecules in the lowest vibrational level \cite{zuchowski2010}. However the energy difference between the entrance and the exit channels of Reaction (\ref{eq:chem-reac}) is so small, that even for molecules identified as {}``non-reactive", Reaction (\ref{eq:chem-reac}) becomes energetically allowed above a small value $v_\mathrm{R}$ of the $\mathrm{AB}$ vibrational quantum number (see Table \ref{tab:v-reac}). Therefore the initial vibrational level can be viewed as a control parameter to study ultracold reactive collisions, and in particular their statistical aspects \cite{mayle2012, gonzalez-martinez2014}.

\begin{table}
\begin{ruledtabular}
\begin{tabular}{ccccc}
         Molecule AB & $\Delta D_0$ (cm$^{-1}$) & $v_\mathrm{R}$ &  $\Delta E_{v_\mathrm{R}}$ (cm$^{-1}$) & Refs. \\
\hline
 $^{23}$Na$^{133}$Cs & 238 & 2 & 392 & \cite{docenko2004a, samuelis2000, amiot2002} \\
  $^{23}$Na$^{87}$Rb &  48 & 1 & 212 & \cite{docenko2004b, samuelis2000, strauss2010} \\
   $^{23}$Na$^{39}$K &  76 & 1 & 246 & \cite{gerdes2008, samuelis2000, falke2008} \\
  $^{39}$K$^{133}$Cs & 238 & 2 & 272 & \cite{ferber2009, falke2008, amiot2002} \\
 $^{87}$Rb$^{133}$Cs &  30 & 1 &  98 & \cite{docenko2011, amiot2002, strauss2010} \\
\end{tabular}
\end{ruledtabular}
\caption{Vibrational quantum number $v_\mathrm{R}$ above which reaction (\ref{eq:chem-reac}) becomes exothermic for the molecules AB which are stable against collisions in their $v=0$ lowest vibrational level (assuming that they are in their lowest rotational level $j=0$). We compare $\Delta D_0=2D_{v=0}(\textrm{AB})-D_{v=0}(\textrm{A}_2)-D_{v=0}(\textrm{B}_2)$, the energy difference between the entrance and the exit channels for molecules in the lowest vibrational level, and $\Delta E_{v_\mathrm{R}}=2\times[E_{v_R}(\textrm{AB})-E_{v=0}(\textrm{AB})]$, twice the energy of the heteronuclear molecule in the vibrational level $v_R$. The difference between these two quantities is the excess energy of reaction (\ref{eq:chem-reac}). The references give the ground state potential energy curves of AB, A$_2$ and B$_2$, that we used to compute the corresponding vibrational spectrum (using the method of Ref. \cite{kokoouline1999}).}
\label{tab:v-reac}
\end{table}

In all these collisions, especially in the {}``universal" regime where AB is assumed to be destroyed with unit probability, the isotropic $C_6$ coefficient characterizing the AB-AB van der Waals interaction $-C_6/R^6$ (where $R$ is the intermolecular distance) turns out to be a crucial parameter \cite{julienne2011, quemener2011a}. Therefore, in this paper we compute the isotropic $C_6$ coefficient between all possible pairs of identical heteronuclear bialkali molecules in the same vibrational level of the electronic ground state $X^1\Sigma^+$. This represents an extension of our previous article \cite{lepers2013}, where we presented the $C_6$ coefficient between molecules in the lowest rovibrational level. We expand the $C_6$ coefficient by distinguishing the contributions from transitions inside and outside the $X$ state, and we show that the hierarchy observed in Ref.~\cite{lepers2013} for $v=0$ persists for a wide range of vibrational levels.

Moreover we discuss the validity of the so-called Moelwyn-Hughes approximation \cite{stone1996}, which aims at calculating in a simple way the $C_6$ coefficient for two molecules in two different vibrational levels. We show that, to work properly, the approximation must be applied separately to each contribution of the $C_6$ coefficient. As the approximation requires the isotropic static dipole polarizability, we give this quantity for all vibrational levels and all heteronuclear bialkali molecules under study. Those calculations are relevant to characterize the collisions between a ground-state and a Feshbach molecule, which are expected to limit the efficiency of the STIRAP transfer to ground state molecules.

The paper is outlined as follows: Section \ref{sec:th} presents the theoretical formalism for molecules in the same vibrational level; in Section \ref{sec:rbcs} we study in details the calculations for the RbCs molecule, focusing on the convergence of our calculations. In section \ref{sec:all-molec} we give our numerical results in a graphical form for all possible pairs of like molecules in the same vibrational level (subsection \ref{sub:same-v}), and we discuss the validity of the Moelwyn-Hughes approximation (subsection \ref{sub:diff-v}). All those results are provided as tables in the supplemental material \cite{supp-mat}. Finally Section \ref{sec:conclu} contains conclusions and prospects. The isotopes used in this article are $^7$Li, $^{23}$Na, $^{39}$K, $^{87}$Rb and $^{133}$Cs; and we expect isotopic substitution to only introduce minor changes in our results. Atomic units of distance (1~$\mathrm{a.u.}\equiv a_0=0.052917721092$~nm)) and energy (1~$\mathrm{a.u.}\equiv 2\times\mathrm{Ryd}=219474.63137$~cm$^{-1}$) are used throughout this paper, except otherwise stated. Occasionally atomic units will also be used for dipole moments (1~a.u.$=2.541 580 59$~D).

\section{Theory \label{sec:th}}

We consider two identical heteronuclear bialkali molecules in the same vibrational level $v$ of the ground electronic state $X^1\Sigma^+$. Since we assume that the molecules are in the lowest rotational level $j_1=j_2=0$, the van der Waals interaction is purely isotropic, and it is characterized by a coefficient $C_6(v)$ which is equal to
\begin{equation}
C_6(v) = \frac{2}{3} \sum_{i'_1 v'_1 i'_2 v'_2}
  \frac{d_{Xv,i'_1v'_1}^2 d_{Xv,i'_2v'_2}^2}
       {\Delta_{Xv,i'_1v'_1}+\Delta_{Xv,i'_2v'_2}}\,,
\label{eq:c6-v}
\end{equation}
where $i'_1$ ($i'_2$) denotes the electronic states of molecule 1 (2) accessible from $X$ through electric-dipole transition, and $v'_1$ ($v'_2$) the corresponding vibrational level. The matrix element $d_{Xv,i'_1v'_1}$ of the dipole moment operator characterizes the strength of those transitions for molecule 1, and similarly for molecule 2.
The prefactor $2/3$ expresses that the only non-vanishing matrix elements correspond to the states $|i'_1v'_1\rangle$ and $|i'_2v'_2\rangle$ with the rotational quantum numbers  $j'_1=j'_2=1$. The labels $j_1, j_2, j'_1, j'_2$ will be most often  omitted in the following for simplicity. In Eq.~(\ref{eq:c6-v}), $\Delta_{Xv,i'_1v'_1}=E_{i'_1,v'_1,1}-E_{Xv0}$ is the energy difference between the rovibronic levels $|i'_1,v'_1,j'_1=1\rangle$ and $|X,v_1,j_1=0\rangle$, and similarly for molecule 2.

We can separate the contributions from each electronic state, namely $C_6(v)=\sum_{i'_1i'_2}C_6^{i'_1i'_2}(v)$ where
\begin{equation}
C_6^{i'_1i'_2}(v) = \frac{2}{3} \sum_{v'_1 v'_2}
  \frac{d_{Xv,i'_1v'_1}^2 d_{Xv,i'_2v'_2}^2}
       {\Delta_{Xv,i'_1v'_1}+\Delta_{Xv,i'_2v'_2}}\,.
\label{eq:c6-i1i2}
\end{equation}
Following our previous work on $v=0$ \cite{lepers2013}, we can define $C_6^\mathrm{g}(v)$, $C_6^\mathrm{e}(v)$, $C_6^\mathrm{g-e}(v)$ and $C_6^\mathrm{e-g}(v)$ which are related to $C_6^{i'_1i'_2}(v)$ as
\begin{eqnarray}
 C_6^\mathrm{g}(v)   & = & C_6^{i'_1=X,i'_2=X}(v) \label{eq:c6g} \\
 C_6^\mathrm{e}(v)   & = & \sum_{i'_1,i'_2\neq X}C_6^{i'_1,i'_2}(v) \label{eq:c6e} \\
 C_6^\mathrm{e-g}(v) \equiv C_6^\mathrm{g-e}(v) & = & \sum_{i'_2\neq X}C_6^{i'_1=X,i'_2}(v) \label{eq:c6ge} \,,
\end{eqnarray}
where for $C_6^\mathrm{g-e}(v)$, we used the symmetry relation $C_6^{i'_2i'_1}(v)=C_6^{i'_1i'_2}(v)$. Note that in Ref.~\cite{lepers2013} addressing the interaction between two identical molecules in their $v=0, j=0$ level, the definition of $C_6^\mathrm{g-e}(v=0)$ quantity corresponds to two times the $C_6^\mathrm{g-e}(v=0)$ coefficient defined in Eq.~(\ref{eq:c6ge}).

Due to the weak variation of the permanent electric dipole moment (PEDM) function with the internuclear distance (see Fig.~\ref{fig:RbCs-PECs} and Ref.~\cite{aymar2005}) the contributions of the vibrational transitions inside the electronic state $X$ are very weak, and $C_6^\mathrm{g}(v)$ can be written to a very good approximation
\begin{equation}
  C_6^\mathrm{g}(v) \approx \frac{d_v^4}{6B_v}\,,
  \label{eq:c6g-v}
\end{equation}
where $d_v$ is the PEDM in the vibrational level $v$ related to the transition $j=0 \rightarrow j'=1$, and $B_v$ the rotational constant of $v$, which is almost equal to half of the energy difference between the ($v, j=0$) and ($v, j'=1$) levels involved in Eq.~(\ref{eq:c6-i1i2}).

\section{Example of the $\textrm{RbCs}$ molecule \label{sec:rbcs}}

In this section we describe in details our calculations, by considering the particular case of RbCs, which has been widely studied in ultracold conditions both experimentally and theoretically \cite{kerman2004a, kerman2004b, sage2005, hudson2008, kotochigova2010a, debatin2011, gabbanini2011, bouloufa2012, takekoshi2012, cho2013, fioretti2013, takekoshi2014,molony2014,shimasaki2015}.

One crucial aspect of our calculation consists in collecting potential energy curves (PECs) and permanent and transition electric dipole moment (PEDM, TEDM) functions. We used up-to-date molecular data extracted from spectroscopic measurements available in the literature, and quantum chemistry calculations from our group otherwise (Table \ref{tab:mol-data}). For instance in RbCs, the PEC of the ground state $X^1\Sigma^+$ was calculated in Ref.~\cite{docenko2011} using the Rydberg-Rees-Klein (RKR) method. The PECs for the states $A^1\Sigma^+$, and $b^3\Pi$ and their spin-orbit interaction are  taken from the spectroscopic investigation of Ref. \cite{docenko2010}. Five additional $^1\Sigma^+$ states, \textit{i.e.} $C$ and (4)--(7), and five $^1\Pi$ states, \textit{i.e.} $B$, $D$ and (3)--(5), were also included. Their PECs as well as the PEDM of the X state and their TEDMs with the X state were calculated in our group with a quantum chemistry approach (see Ref. \cite{aymar2005} for details). If necessary, the experimental or computed PECs are matched to an asymptotic long-range expansion following refs \cite{docenko2011,marinescu1999}. The results for the lowest electronic states are displayed in Fig.~\ref{fig:RbCs-PECs}. For every electronic state, the vibrational continuum is taken into account. The energies and wave functions of the bound and free levels are calculated using our mapped Fourier grid Hamiltonian method \cite{kokoouline1999}. 

For a proper convergence of the calculations, the excitation of the molecular core states must be included in Eqs.~(\ref{eq:c6e}) and (\ref{eq:c6ge}), beside the one of the molecular valence states. As such core states are unknown and quite difficult to evaluate, we relied on the fact that their excitation energy is much larger than the one of the valence state, so that we took their contributions into account through the excitations of both ionic cores of the diatomic molecule. The procedure is described in details in Refs.~\cite{lepers2011a,vexiau_phd_2012,vexiau2015} and is not repeated here. These contributions are labeled as partial coefficients $C_6^{i' c}$ where the index $c$ refers collectively to core excitations.

In the rest of this section we will discuss the influence of the different electronic states, and of the inclusion of the vibrational continua. 

We start by giving our computed values of $C_6^\mathrm{g}(v)$, $C_6^\mathrm{e}(v)$ and $C_6^\mathrm{g-e}(v)$ as functions of the vibrational quantum number $v$ (see Fig.~\ref{fig:C6_contrib} and supplemental material \cite{supp-mat}). The most remarkable trend is the strong decrease in the $C_6$ coefficient as a function of $v$, which follows the strong decrease in $C_6^\mathrm{g}$. This is due to the variation of the PEDM, which is maximal in the vicinity of the equilibrium distance, and which diminishes at large distances, and thus for high $v$. Figure~\ref{fig:C6_contrib} also shows a slight enhancement of $C_6^{e}$ which, together with the decrease in $C_6^{g}$, induces a maximum $C_6$ of $1.494\times 10^5$ a.u. for $v=14$.

Near the dissociation limit, the PEDM and correspondingly $C_6^\mathrm{g}(v)$ tend to zero. The only contribution comes from electronically-excited states. The two RbCs molecules almost behave like four free atoms, and the $C_6$ coefficient can be written
\begin{equation}
  C_6(v\to\infty) = C_6^\textrm{Rb-Rb} + C_6^\textrm{Cs-Cs}
                  + 2C_6^\textrm{Rb-Cs}.
\label{eq:C6-disso}
\end{equation}
Equation (\ref{eq:C6-disso}) is actually a very good approximation, since our calculations give 22510 a.u., while we obtain 22862 a.u. by taking the atom-atom coefficient of Ref.~\cite{derevianko2010}.

Now we examine specifically the contribution $C_6^\mathrm{e}$ of the electronically-excited states. On Fig.~\ref{fig:C6_part} we present the partial coefficients $C_6^{i'_1i'_2}$ as functions of $v$, for the states $i'_1$, $i'_2=A$, $B$, $C$ and $D$. We also include the contribution from the excitation of the ionic core states, which we suppose to be independent from $v$. [Note that the crossed contributions $i'_1\neq i'_2$ have to be counted twice to get the total $C_6^{e}(v)$.] The coefficients $C_6^{i'_1i'_2}$ vary significantly under the combined effects of the TEDM variations and Franck-Condon factors (FCFs). For example, $C_6^{BB}$ is maximum for $v=0$ since the equilibrium distances of $X$ and $B$ states are very close to each other. Then $C_6^{BB}$ decreases with $v$ due to a significantly worse FCF. The latter is still poor in the high-$v$ region, but $C_6^{BB}$ follows the enhancement of the $X\to B$ TDM. On the contrary the contributions $C_6^{Ci'_2}$ coming from the $C^1\Sigma^+$ state are smaller, since the minimum of $X$ and $C$ states are noticeably shifted, inducing poor FCFs.

On Figs.~\ref{fig:C6_trans} and \ref{fig:C6_perm} we discuss the convergence of the sum-over-state formula (\ref{eq:c6-i1i2}) with respect to the number of included electronically-excited states. Beyond the four first excited states ($A$, $B$, $C$ and $D$) each additional state increases the $C_6^{e}$ value by about 150 a.u., that is a little less than $1\%$. This gives a total increase of $C_6^\mathrm{e}(v)$ by about 1000 a.u. ($5\%$) when we include up to the $(7)^1\Sigma^+$ and the $(5)^1\Pi$ states. This increase only represents less than $1\%$ of the total $C_6$ for $v=0$ (see Fig.~\ref{fig:C6_perm}). To illustrate the effects of higher states which are not included in our computed values above (see Fig.~\ref{fig:C6_contrib}), we can mention that the $(8)^1\Pi$ adds only $0.05\%$ (8 a.u.). Moreover  an estimate for the other excited states, which have a TEDM with $X$ even smaller than $(8)^1\Pi$, reveals that these states bring a contribution at least 4 times weaker than the state $(8)^1\Pi$.

We have also examined the sensitivity of our results due to the PECs that we use as input data, in particular for the $A^1\Sigma^+$ state. As said above, the calculations of Fig.~\ref{fig:C6_contrib.eps} are performed with a RKR curve, and the spin-orbit coupling with the $b^3\Pi$ state is also taken into account. The $C_6$ coefficients can be evaluated taking instead the $A^1\Sigma^+$ RKR curve without any spin-orbit coupling, or a PEC from a quantum chemistry calculation with or without the short-range repulsive term taken from \cite{jeung1997}. As shown on Fig.~\ref{fig:C6_abinitio}, the final results are weakly sensitive to the details of the PECs. Even if the different PECs support a different total number of vibrational levels, the sum-over-state formula (\ref{eq:c6-v1v2}), which does not favor any level with respect to its neighbors, washes out such differences. The conclusion in this respect would drastically change if we calculate the dynamic dipole polarizability for frequencies close to the transition energies \cite{vexiau2015}.

Now we check the influence of the vibrational continua of the electronically-excited states on the partial (Fig.~\ref{fig:C6_continuum-part}) and total (Fig.~\ref{fig:C6_continuum}) $C_6$ coefficients. For each excited state, the influence of the continuum is negligible for the lowest $v$ levels, quite abruptly increases for intermediate ones  reaching up to 50 \% of the total $C_6$ value or beyond, and drops down for the very last vibrational levels. Since the TEDMs are smoothly varying with the distance, these dramatic variations are related to the FCFs. For small $v$ of the electronic ground state, the wave function is so localized around the equilibrium distance, that the FCF with excited-states continuum wave functions is vanishingly small. The FCF is enhanced for moderately-excited vibrational levels, because a significant part of the corresponding wave functions is located around their inner turning point, where the overlap with continuum wave functions is favorable. On the contrary, for the highest $v$ the wave functions are mostly located around their outer turning point where the best overlap occurs with the bound vibrational levels of the electronically-excited states. 

Finally, the last limiting factor for the precision of our calculation is the accuracy of the TEDMs which cannot be easily measured. The comparison of computed TEDMs among various methods for the transitions involving the lowest electronic states has been discussed for instance in one of our previous work \cite{aymar2007}, suggesting that the observed typical difference of about $3\%$ is representative of the TEDMs uncertainty. For the molecules where $C_6^g$ is dominant, the comparison of the calculated PEDM for the $v=0$ level \cite{aymar2005,supp-mat} with reported experimental values is also noteworthy. A recent measurement for RbCs \cite{molony2014} reports $d_{v=0}=1.225(3)(8)$~D, compared to the present computed one, $d_{v=0}=1.246$~D; a similar agreement is observed for NaK calculated at 2.783~D, compared to an experimental value of 2.72(6)~D \cite{gerdes2011}. It is thus difficult to yield a proper value of the uncertainty of the computed $C_6$ coefficients, but we conservatively estimate it to about 15\%.

\section{Results for all heteronuclear bialkali molecules  
\label{sec:all-molec}}

\subsection{Molecules in the same vibrational level 
\label{sub:same-v}}

The same method has been applied to all the other heteronuclear bialkali molecules. The input data used for the calculations, especially the PECs, are presented in Table \ref{tab:mol-data}. In Ref.~\cite{lepers2013} we already gave our results for $v=0$ and we compared them with the literature \cite{kotochigova2010a, quemener2011a, buchachenko2012, byrd2012b, zuchowski2013}. But to our knowledge the present article is the first study concerning vibrationally-excited levels. The results are given on Figs. \ref{fig:C6_binding} and \ref{fig:C6_total} and in the supplemental material \cite{supp-mat}. In order to provide graphs with all the molecules, we plotted the partial $C_6^\mathrm{e}$ and total $C_6$ coefficients as functions of the binding energy. We see very similar trends with respect to the binding energy, hence to the vibrational quantum number $v$. The coefficient $C_6^\mathrm{e}$ increases with $v$ except near the dissociation limit (see Fig.~\ref{fig:C6_binding}), while the total $C_6$ coefficient strongly decreases with $v$ except for LiNa and KRb (see Fig.~\ref{fig:C6_total}). This is due to the same reason than the one invoked in Ref. \cite{lepers2013}. For LiNa and KRb, the contribution $C_6^\mathrm{e}$ dominates over $C_6^\mathrm{g}$ due to the weak PDM; on the contrary for the other molecules, although $C_6^\mathrm{g}$ shrinks following the decrease of the PEDM with $v$ (see Eq.~(\ref{eq:c6g-v})), it remains the largest contribution except for the very last vibrational levels.

\subsection{Molecules in different vibrational levels \label{sub:diff-v}}

Now we turn to the calculation of the coefficient $C_6(v_1,v_2)$ describing the interaction between two ground-state molecules in two different vibrational levels $v_1$ and $v_2$ (still with $j_1=j_2=0$), either close to or far from each other. This is for instance relevant for the interaction between a ground-state and a Feshbach molecule. The coefficient $C_6(v_1,v_2)$ reads
\begin{equation}
C_6(v_1,v_2) = \frac{2}{3} \sum_{i'_1 v'_1 i'_2 v'_2}
  \frac{d_{Xv_1,i'_1v'_1}^2 d_{Xv_2,i'_2v'_2}^2}
       {\Delta_{Xv_1,i'_1v'_1}+\Delta_{Xv_2,i'_2v'_2}}\,,
\label{eq:c6-v1v2}
\end{equation}
where $\Delta_{Xv_k,i'_kv'_k}=E_{i'_k,v'_k,j'_k=1}-E_{X,v_k,j_k=0}$ and $d_{Xv_k,i'_kv'_k}$ are respectively the transition energies and transition dipole moments for molecules $k=1,2$.

Giving a numerical value $C_6(v_1,v_2)$ for each couple ($v_1,v_2$) and for each molecule would be particularly cumbersome. Therefore it is appropriate to express the coefficient $C_6(v_1,v_2)$ as a function of the coefficients $C_6(v_1)$ and $C_6(v_2)$ between two identical levels, using the Moelwyn-Hughes (MH) approximation \cite{stone1996}
\begin{eqnarray}
  \left(\frac{C_6^\mathrm{MH}(v_1,v_2)}{2}\right)^{-1} 
 & = & \left(\frac{\alpha(v_2)}{\alpha(v_1)}C_6(v_1)\right)^{-1}
  \nonumber \\
 & + & \left(\frac{\alpha(v_1)}{\alpha(v_2)}C_6(v_2)\right)^{-1} ,
\label{eq:c6-v1v2-MH}
\end{eqnarray}
where $\alpha(v)$ is the static dipole polarizability in the vibrational level $v$. For polar molecules $\alpha(v)$ can be expanded as
\begin{equation}
  \alpha(v) = \alpha_\mathrm{g}(v) + \alpha_\mathrm{e}(v) ,
\label{eq:alpha-v}
\end{equation}
where $\alpha_\mathrm{g}(v)$ and $\alpha_\mathrm{e}(v)$ respectively denote the contributions from the transitions inside the ground state $X$, and to the electronically-excited states. In analogy to $C_6^\mathrm{g}(v)$, $\alpha_\mathrm{g}(v)$ is dominated by the strong purely rotational transition, and thus can be written, to a very good approximation,
\begin{equation}
  \alpha_\mathrm{g}(v) = \frac{d_v^2}{3B_v}\,.
  \label{eq:alpha-g}
\end{equation}
In order to discuss the validity of the MH approximation (\ref{eq:c6-v1v2-MH}), we assume that the contributions from the electronically-excited states can be reduced to a single effective transition towards $(i^*,v^*)$, which means 
\begin{equation}
  \alpha_\mathrm{e}(v) = \frac{2d_{Xv,i^*v^*}^2}
                              {3\Delta_{Xv,i^*v^*}}
  \label{eq:alpha-e-eff}
\end{equation}
and
\begin{eqnarray}
  C_6^\mathrm{e}(v) & = & \frac{d_{Xv,i^*v^*}^4}
                               {3\Delta_{Xv,i^*v^*}} \,,
  \label{eq:c6e-v-eff} \\
  C_6^\mathrm{e-g} \equiv C_6^\mathrm{g-e}(v) & = & \frac{2d_v^2 d_{Xv,i^*v^*}^2}
                                 {3(2B_v+\Delta_{Xv,i^*v^*})}\,.
  \label{eq:c6ge-v-eff}
\end{eqnarray}

Now we can compare the $C_6$ coefficients between two different vibrational levels $v_1$ and $v_2$, calculated either directly or with the MH approximation. The direct calculation gives for each contribution
\begin{eqnarray}
 C_6(v_1,v_2)            & = & C_6^\mathrm{g}(v_1,v_2)
                             + C_6^\mathrm{e}(v_1,v_2) \nonumber \\
                         & + & C_6^\mathrm{g-e}(v_1,v_2)
                             + C_6^\mathrm{e-g}(v_1,v_2)
 \label{eq:c6-v1v2-contrib} \\
 C_6^\mathrm{g}(v_1,v_2) & = & \frac{d_{v_1}^2 d_{v_2}^2}
                                       {3(B_{v_1}+B_{v_2})}
 \label{eq:c6g-v1v2} \\
 C_6^\mathrm{e}(v_1,v_2) & = & \frac{2d_{Xv_1,i_1^*v_1^*}^2 
                                      d_{Xv_2,i_2^*v_2^*}^2}
                                    {3(\Delta_{Xv_1,i_1^*v_1^*}
                                      +\Delta_{Xv_2,i_2^*v_2^*})} 
 \label{eq:c6e-v1v2} \\
 C_6^\mathrm{g-e}(v_1,v_2) & = & \frac{2d_{v_1}^2 
                                       d_{Xv_2,i_2^*v_2^*}^2}
                                      {3(2B_{v_1}
                                        +\Delta_{Xv_2,i_2^*v_2^*})}  
 \label{eq:c6ge-v1v2} \\
 C_6^\mathrm{e-g}(v_1,v_2) & = & \frac{2d_{Xv_1,i_1^*v_1^*}^2 
                                        d_{v_2}^2}
                                      {3(\Delta_{Xv_1,i_1^*v_1^*}
                                        +2B_{v_2})} \,,
 \label{eq:c6eg-v1v2}
\end{eqnarray}
Note that Eqs.~(\ref{eq:c6ge-v1v2}) and (\ref{eq:c6eg-v1v2}) are not equivalent when $v_1\neq v_2$.
In order to use the MH approximation, we need to apply Equation (\ref{eq:c6-v1v2-MH}) using the following ingredients: $\alpha(v_k)$ is obtained by applying Eqs.~(\ref{eq:alpha-v})--(\ref{eq:alpha-e-eff}) with $v\equiv v_k$ ($k=1,2$); and $C_6(v_k)$ is obtained by adding Eqs.~(\ref{eq:c6g-v}), (\ref{eq:c6e-v-eff}) and (\ref{eq:c6ge-v-eff}) for $v\equiv v_k$ ($k=1,2$).
It is then straightforward to see that the value of $C_6^\mathrm{MH}(v_1,v_2)$ obtained in this way cannot be equal to $C_6(v_1,v_2)$ given in Eq.~(\ref{eq:c6-v1v2-contrib}). However, to go beyond this general statement, it is worthwhile to look closely at some particular cases.

For weakly polar molecules like LiNa or KRb in the ground vibrational level, $C_6^\mathrm{g}(v=0)$ and $C_6^\mathrm{e}(v=0)$ are comparable, although the ground-state polarizability is dominant, $\alpha_\mathrm{g}(v=0)\gg\alpha_\mathrm{e}(v=0)$ (see respectively Tab.~II and Fig.~1 of Ref.~\cite{lepers2013}). If we compare Eqs.~(\ref{eq:c6-v1v2}) and (\ref{eq:c6-v1v2-MH}), by neglecting $\alpha_\mathrm{e}$ (see Eq.~(\ref{eq:alpha-e-eff})), we see after some calculations that the MH approximation is not valid.
On the contrary we can show that it works properly either for strongly polar or vibrationally highly-excited molecules. In the former case, which corresponds to all the molecules in low-lying vibrational levels except LiNa and KRb, the purely rotational transition is dominant, \textit{i.e.}~$\alpha_\mathrm{g}(v_k)\gg\alpha_\mathrm{e}(v_k)$ and $C_6^\mathrm{g}(v_k)\gg C_6^\mathrm{e}(v_k),\,C_6^\mathrm{g-e}(v_k)$ for $v_k=1$ and 2, and so we obtain $C_6(v_1,v_2)\approx C_6^\mathrm{MH}(v_1,v_2)\approx d_{v_1}^2 d_{v_2}^2 /  3(B_{v_1}+B_{v_2})$.
In the latter case, the PDM $d_{v_k}$, as well as $\alpha_\mathrm{g}(v_k)$ and $C_6^\mathrm{g}(v_k)$, tend to zero, and so we obtain $C_6(v_1,v_2)\approx C_6^\mathrm{MH}(v_1,v_2)\approx 2d_{Xv_1,i_1^*v_1^*}^2 d_{Xv_2,i_2^*v_2^*}^2 / 3(\Delta_{Xv_1,i_1^*v_1^*}+\Delta_{Xv_2,i_2^*v_2^*})$.

Those analytical estimates suggest that it is more appropriate to apply the MH approximation (\ref{eq:c6-v1v2-MH}) to the partial coefficients $C_6^\mathrm{g}(v_1,v_2)$, $C_6^\mathrm{e}(v_1,v_2)$ $C_6^\mathrm{g-e}(v_1,v_2)$ and $C_6^\mathrm{e-g
}(v_1,v_2)$ separately, rather than to the total coefficient $C_6(v_1,v_2)$, namely
\begin{eqnarray}
  \left(\frac{C_6^\mathrm{g,MH}(v_1,v_2)}{2}\right)^{-1} 
 & = & \left(\frac{\alpha_\mathrm{g}(v_2)}{\alpha_\mathrm{g}(v_1)}C_6^\mathrm{g}(v_1)\right)^{-1}
  \nonumber \\
 & + & \left(\frac{\alpha_\mathrm{g}(v_1)}{\alpha_\mathrm{g}(v_2)}C_6^\mathrm{g}(v_2)\right)^{-1} 
\label{eq:c6-v1v2-MH-g} \\
  \left(\frac{C_6^\mathrm{e,MH}(v_1,v_2)}{2}\right)^{-1} 
 & = & \left(\frac{\alpha_\mathrm{e}(v_2)}{\alpha_\mathrm{e}(v_1)}C_6^\mathrm{e}(v_1)\right)^{-1}
  \nonumber \\
 & + & \left(\frac{\alpha_\mathrm{e}(v_1)}{\alpha_\mathrm{e}(v_2)}C_6^\mathrm{e}(v_2)\right)^{-1} 
\label{eq:c6-v1v2-MH-e} \\
  \left(\frac{C_6^\mathrm{g-e,MH}(v_1,v_2)}{2}\right)^{-1} 
 & = & \left(\frac{\alpha_\mathrm{e}(v_2)}{\alpha_\mathrm{g}(v_1)}C_6^\mathrm{g}(v_1)\right)^{-1}
  \nonumber \\
 & + & \left(\frac{\alpha_\mathrm{g}(v_1)}{\alpha_\mathrm{e}(v_2)}C_6^\mathrm{e}(v_2)\right)^{-1} 
\label{eq:c6-v1v2-MH-g-e} \\
  \left(\frac{C_6^\mathrm{e-g,MH}(v_1,v_2)}{2}\right)^{-1} 
 & = & \left(\frac{\alpha_\mathrm{g}(v_2)}{\alpha_\mathrm{e}(v_1)}C_6^\mathrm{e}(v_1)\right)^{-1}
  \nonumber \\
 & + & \left(\frac{\alpha_\mathrm{e}(v_1)}{\alpha_\mathrm{g}(v_2)}C_6^\mathrm{g}(v_2)\right)^{-1} 
\label{eq:c6-v1v2-MH-e-g}
\end{eqnarray}
This is confirmed by the results of Fig.~\ref{fig:C6_MH}, which displays the total $C_6$ coefficient between two RbCs molecules, one being in the ground rovronic level, and the other in an arbitrary vibrational level $v$. As long as the $C_6^\mathrm{g}$ contribution is dominant, we see that is sufficient to apply the MH approximation to the total $C_6$ coefficient. But when the contribution of the electronically-excited states become significant, \textit{i.e.} from $v\approx80$ (see Fig.~\ref{fig:C6_contrib}), it is necessary to apply the MH approximation to each contribution separately. This requires to know the coefficients $C_6^\mathrm{g}(v_k)$, $C_6^\mathrm{e}(v_k)$, and the static dipole polarizabilities $\alpha_\mathrm{g}(v_k)$ and $\alpha_\mathrm{e}(v_k)$, for the two levels of interest $v_1$ and $v_2$. To that end we give in the supplemental material those five quantities for all vibrational levels $v$ and all molecules \cite{supp-mat}.

\section{Conclusions and outlook \label{sec:conclu}}

In this article, we have computed the isotropic $C_6(v)$ coefficient between all pairs of identical alkali-metal diatomic molecules made of $^7$Li, $^{23}$Na, $^{39}$K, $^{87}$Rb and $^{133}$Cs, lying in the same vibrational level $v$ of the electronic ground state $X^1\Sigma^+$. Following our previous work \cite{lepers2013} we have expanded $C_6(v)$ as a sum of three contributions, $C_6^\mathrm{g}(v)$, $C_6^\mathrm{e}(v)$ and $C_6^\mathrm{g-e}(v)$, coming respectively from transitions inside the electronic ground state, from transitions to electronically-excited states, and from crossed terms.
To a very good approximation, $C_6^\mathrm{g}(v)$ is dominated by the purely rotational transition $|X,v_k=v,j_k=0\rangle\to|X,v'_k=v,j'_k=1\rangle$ ($k=1,2$), and so following the permanent electric dipole moment $d_v$, it strongly decreases with the vibrational quantum number $v$. In comparison the variation is $C_6^\mathrm{e}(v)$ is much smoother, and $C_6^\mathrm{g-e}(v)$ is always vanishingly small \cite{supp-mat}.

We observe that the hierarchy between contributions that we established in our previous paper for $v=0$ is observable for a wide range of vibrational quantum numbers $v$. For all molecules but LiNa and KRb, $C_6^\mathrm{g}(v)$ is very strong, and so we expect the effect of mutual orientation described in \cite{lepers2013} to occur below the intermolecular distance $R^*_v=(d_v^2/B_v)^{1/3}$. In all other cases (LiNa, KRb and other molecules close to the dissociation limit), $C_6^\mathrm{e}(v)$ is dominant, and there is no mutual orientation.

In order to characterize the van der Waals interaction between molecules in different vibrational levels $v_1$ and $v_2$, we also discuss the validity of the Moelwyn-Hughes approximation, in which the coefficient $C_6(v_1,v_2)$ is expressed as a function of $C_6(v_1)$ and $C_6(v_2)$. We show that it is more appropriate to apply the approximation to the partial coefficients $C_6^\mathrm{g}(v_1,v_2)$, $C_6^\mathrm{e}(v_1,v_2)$, $C_6^\mathrm{g-e}(v_1,v_2)$ and $C_6^\mathrm{e-g}(v_1,v_2)$, than to the total one $C_6(v_1,v_2)$. Such conclusions can be extended to the interaction between two (different) molecules in arbitrary rovibrational levels. In particular for rotationally-excited levels, anisotropic $C_6^\mathrm{e}$, $C_6^\mathrm{g-e}$ and $C_6^\mathrm{e-g}$ coefficients would come into play, which could also be calculated with the Moelwyn-Hughes approximation. Another readily accessible extension concerns the long-range interaction between a molecule in their lowest $v=0, j=0$ level, and a Feshbach molecule. Indeed we have shown \cite{vexiau2011} that the dipole polarizability of a diatomic molecule in a weakly-bound Feshbach level is well approximated by the one of the uppermost level of a single potential curve, itself very close to the sum of the two individual dipole polarizabilities \cite{supp-mat}.

\appendix

\section{Molecular data}

In this appendix we specify the input data -- potential-energy curves, permanent and transition dipole moments and spin-orbit couplings -- that we used in our calculations. In Table \ref{tab:mol-data} we give in particular the references for available experimental data. All the quantum-chemical data that we used were computed in our group \cite{aymar2005,vexiau2015}.

\begingroup
\squeezetable
\begin{table*}[!p]
	
        \begin{tabularx}{\textwidth}{C{0.08}C{0.08}C{0.16}C{0.08}C{0.16}C{0.08}C{0.08}L{0.24}}
        \hline
		Molecule & {$X^1\Sigma^+$\linebreak experimental state} & Experimental excited states & Long Range & \ab PECs \cite{vexiau2015} & { SOCME} & 
		{ PDMs\linebreak TDMs} & 
		{\centering Notes}\tabularnewline
		\hline
		KCs & \cite{ferber2009} & $A^1\Sigma^+$, $b^3\Pi$ \cite{kruzins2010}, $B^1\Pi$ \cite{birzniece2012}, 
			$E^1\Sigma^+$ \cite{busevica2011} & \cite{kruzins2010,busevica2011,marinescu1999a} & 
			$(3,5,6)^1\Sigma^+$, $(2,3)^1\Pi$ & 
			$(b/A)$ \cite{kruzins2010} & 
			\cite{aymar2005}\linebreak  \cite{vexiau2015} & 
			$B^1\Pi$ : RKR + \ab + long range.\linebreak
			$E^1\Sigma^+$ : modified empirical PEC to converge toward the unperturbed atomic limit (4s+5d)\tabularnewline
		\hline
		KRb & \cite{pashov2007} & $A^1\Sigma^+$ \cite{kim2012}, $b^3\Pi$ \cite{kim2011}, 
			$C^1\Sigma^+$ \cite{amiot1999}, $B^1\Pi$ \cite{kasahara1999}, $D^1\Pi$ \cite{amiot2000a}, $(3)^1\Pi$ \cite{amiot1999}& 
			\cite{marinescu1999a} & $(4,5)^1\Sigma^+$ & $(b/A)$ \cite{docenko2007} & 
			\cite{aymar2005}\linebreak  \cite{vexiau2015} &
			$A^1\Sigma^+$, $b^3\Pi$ : spectroscopic data on a very limited range $\Rightarrow$ shift of \ab PECs to fit with data. 
			SOCME from NaRb is used.\linebreak
			$(1,2)^1\Pi$, $C^1\Sigma^+$  : RKR + \ab + long range\linebreak
			$(3)^1\Pi$ : \ab + experimental $T_e$
			\tabularnewline
		\hline
		RbCs & \cite{docenko2011} & $A^1\Sigma^+$, $b^3\Pi$ \cite{docenko2010} & \cite{docenko2011,marinescu1999a} & 
			$(3-7)^1\Sigma^+$, $(1-5)^1\Pi$ & $(b/A)$ \cite{docenko2010} & \cite{aymar2005}\linebreak  \cite{vexiau2015} &  \tabularnewline
		\hline
		LiNa & \cite{fellows1991} & $A^1\Sigma^+$ \cite{fellows1989}, $C^1\Sigma^+$ \cite{fellows1990}, 
			$E^1\Sigma^+$ \cite{bang2005} & \cite{fellows1991,marinescu1999a} & $(5)^1\Sigma^+$, $(1,2,3)^1\Pi$ & -- & 
			\cite{aymar2005}\linebreak  \cite{vexiau2015} &
			Weak SO interaction $\Rightarrow$ not included in the calculations.\tabularnewline
		\hline
		LiK & \cite{tiemann2009} & $A^1\Sigma^+$ \cite{grochola2012}, $C^1\Sigma^+$ \cite{jastrzebski2001}, 
			$B^1\Pi$ \cite{jastrzebski2001} & \cite{tiemann2009,marinescu1999a} & 
			$(4,5)^1\Sigma^+$, $(2,3)^1\Pi$ & $(b/A)$ \cite{docenko2007} &
			\cite{aymar2005}\linebreak  \cite{vexiau2015} &
			Rescaled SOCME from NaRb is used.\linebreak
			$A^1\Sigma^+$, $C^1\Sigma^+$, $B^1\Pi$ : RKR + \ab + long range\tabularnewline
		\hline
		LiRb & \cite{ivanova2011} &  $B^1\Pi$ \cite{ivanova2013}, $C^1\Sigma^+$ \cite{ivanova2013}, $D^1\Pi$ \cite{ivanova2013}  & \cite{ivanova2011,ivanova2013,marinescu1999a} & 
			$(2,4,5)^1\Sigma^+$, $(3)^1\Pi$ & $(b/A)$ \cite{docenko2007} & 
			\cite{aymar2005}\linebreak  \cite{vexiau2015} & SOCME from NaRb is used \tabularnewline
		\hline
		LiCs & \cite{staanum2007} & $B^1\Pi$ \cite{grochola2009} & \cite{staanum2007,grochola2009,marinescu1999a} & 
			$(2-5)^1\Sigma^+$, $(2,3)^1\Pi$ & $(b/A)$ \cite{zaharova2009} &
			\cite{aymar2005}\linebreak  \cite{vexiau2015} & SOCME from NaCs is used \tabularnewline
		\hline
		NaK & \cite{gerdes2008} & $b^3\Pi$ \cite{ross1986a}, $B^1\Pi$ \cite{kasahara1991}, $C^1\Sigma^+$ \cite{ross2004}, 
			$D^1\Pi$, $d^3\Pi$ \cite{adohikrou2008} & \cite{gerdes2008,adohikrou2008,marinescu1999a} & 
			$(2,4,5)^1\Sigma^+$, $(2-4)^1\Pi$ & $(d/D)$ \cite{adohikrou2008}, $(b/A)$ \cite{docenko2007} & 
			\cite{aymar2007} &
			Rescaled SOCME from NaRb for $(b/A)$ is used (qualitative agreement $6\%$ with \textit{ab initio} from \cite{manaa1999})\tabularnewline
		\hline
		NaRb & \cite{docenko2004b} & $A^1\Sigma^+$, $b^3\Pi$ \cite{docenko2007}, $B^1\Pi$ \cite{pashov2006}, 
			$C^1\Sigma^+$ \cite{jastrzebski2005}, $D^1\Pi$ \cite{docenko2005},$(4)^1\Pi$\cite{bang2009} & 
			\cite{docenko2004b,docenko2007,pashov2006,jastrzebski2005,docenko2005,marinescu1999a} & 
			$(4-5)^1\Sigma^+$, $(3,5)^1\Pi$ & $(b/A)$ \cite{docenko2007} & \cite{aymar2007} & \tabularnewline
		\hline
		NaCs & \cite{docenko2004a} & $A^1\Sigma^+$, $b^3\Pi$ \cite{zaharova2009}, $B^1\Pi$ \cite{zaharova2007}, 
			$(3)^1\Pi$ \cite{docenko2006a} & \cite{docenko2004a,zaharova2007,docenko2006a,marinescu1999a} & 
			$(3,4,5)^1\Sigma^+$, $(2,4,5)^1\Pi$ & $(b/A)$ \cite{zaharova2009} & \cite{aymar2007} & \tabularnewline
		\hline	
	\end{tabularx}
\caption{\label{tab:mol-data} References for all the electronic PECs used in the calculations (SOCME $\equiv$ Spin-Orbit Coupling Matrix Elements). For heavy molecules (RbCs, KRb, KCs) we have also added the short range core-core repulsion from \cite{jeung1997} to the \textit{ab initio} PECs. This term is less important for lighter molecules. For the \textit{ab initio} PECs of LiRb and NaRb we have extrapolated the short-range potential by comparing our ground state \textit{ab initio} PEC to the RKR potential, for the other molecules the term is not included.}
\end{table*}

\section*{Acknowledgements}

We thank Demis Borsalino for the preparation of Table \ref{tab:mol-data}. R.V. acknowledges partial support from Agence Nationale de la Recherche (ANR), under the project COPOMOL (contract ANR-13-IS04-0004-01).


%

\newpage
\begin{figure*}
\begin{center}
\includegraphics[width=8cm]{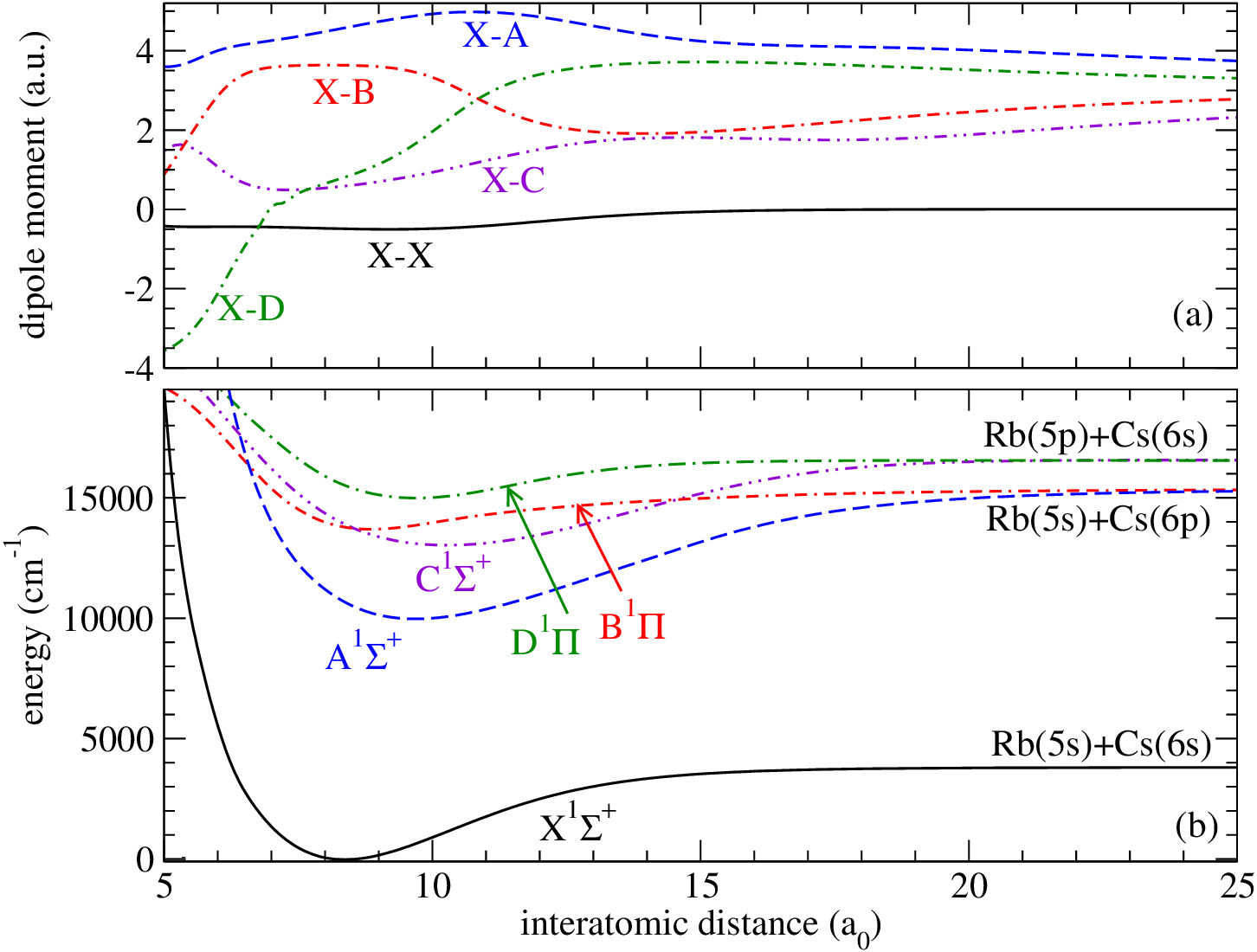}
\caption{\label{fig:RbCs-PECs} (Color online) Selected permanent and transition dipole moments (a) and potential-energy curves  (b) for the lowest singlet states of RbCs: $X^1\Sigma^+$ (solid line), $A^1\Sigma^+$ (dashed line), $B^1\Pi$ (dash-dash-dotted line), $C^1\Sigma^+$ (dash-dot-dotted line), $D^1\Pi$ (dash-dotted line). The curve labeled X-X corresponds to the permanent dipole moment of the $X^1\Sigma^+$ state, and the others to transition dipole moments from the X state. } 
\end{center}
\end{figure*}

\newpage
\begin{figure*}
\begin{center}
\includegraphics[width=8cm]{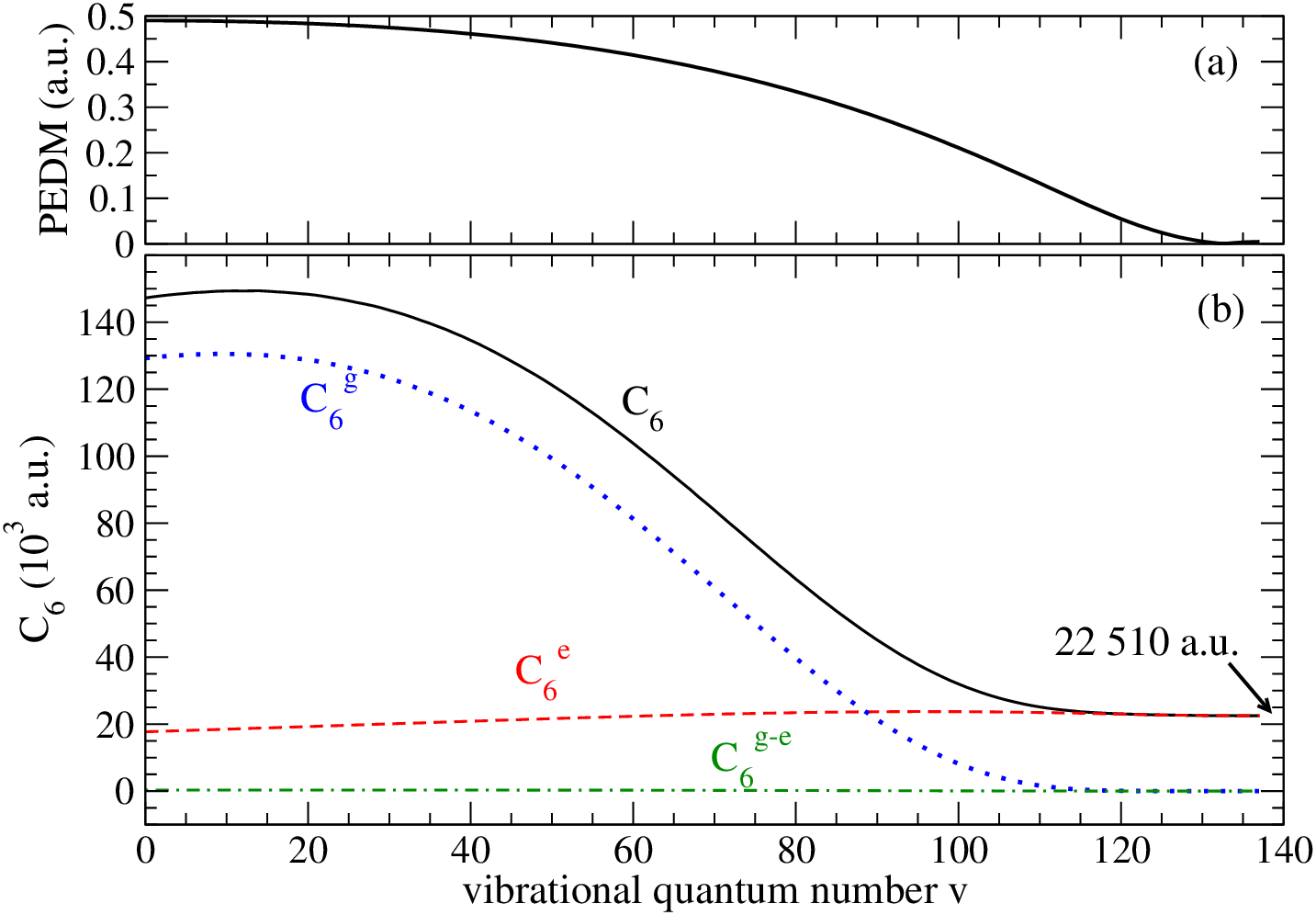}
\caption{\label{fig:C6_contrib} (Color online) (a) Permanent electric dipole moment $d_v$ of the $^{87}$Rb$^{133}$Cs vibrational levels $v$ in the frame fixed to the molecule (see Eq.~(\ref{eq:c6g-v})). (b) Total $C_6$ coefficient (solid lines) as well as $C_6^\mathrm{g}$ (dotted lines), $C_6^\mathrm{e}$ (dashed lines) and $C_6^\mathrm{g-e}$ (dash-dotted lines) contributions for the interaction between two RbCs $|X^1\Sigma^+,v,j=0\rangle$ molecules as functions of the vibrational quantum number $v$. The $C_6$ value indicated by the arrow corresponds to the last vibrational level $v=137$.} 
\end{center}
\end{figure*}

\newpage
\begin{figure*}
\begin{center}
\includegraphics[width=8cm]{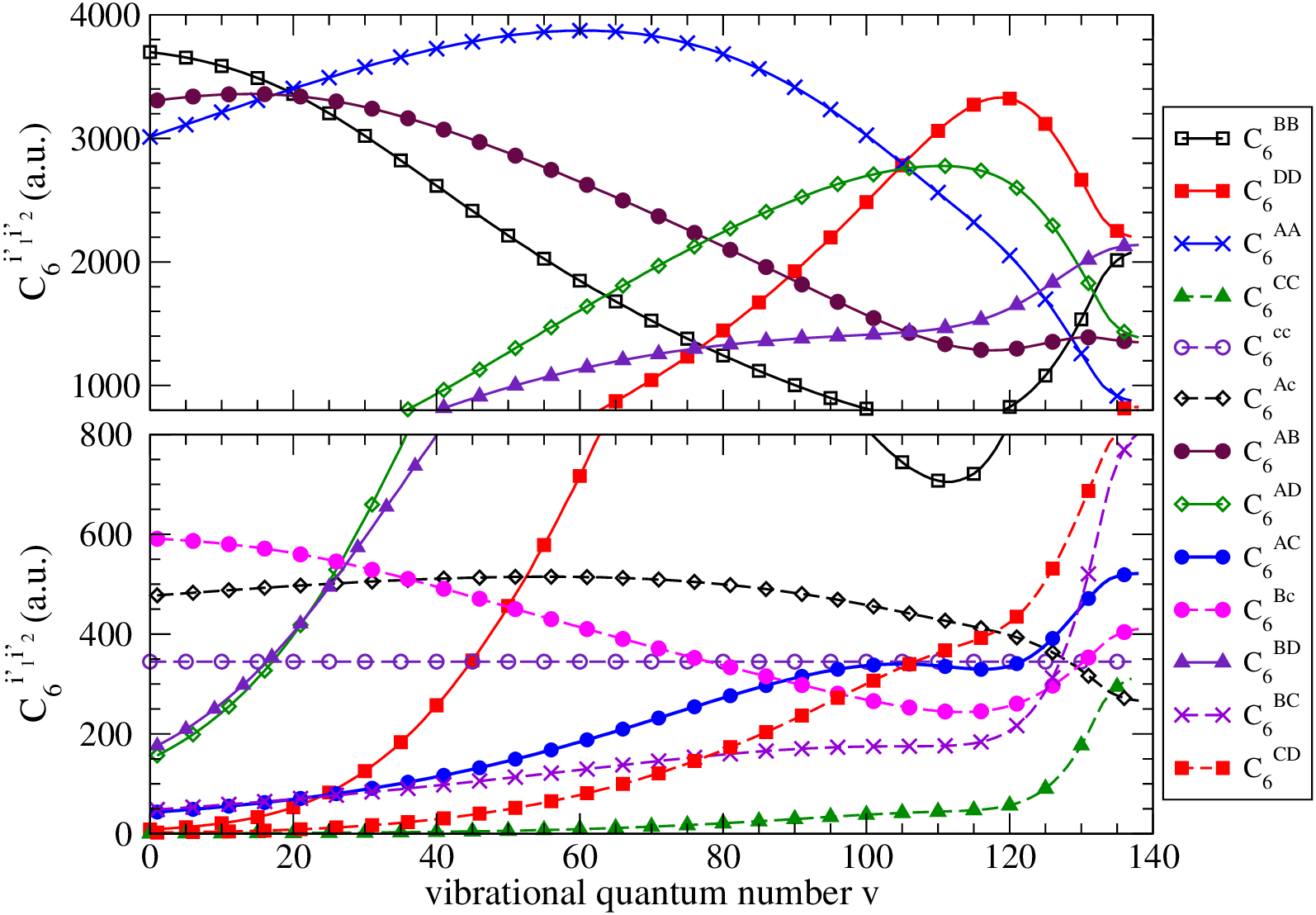}
\caption{\label{fig:C6_part} (Color online) Main partial coefficients $C_6^{i'_1i'_2}(v)$ (see Eq.~(\ref{eq:c6-i1i2})) as functions of the vibrational quantum number $v$, related to the electronically-excited states $A^1\Sigma^+$ (including spin-orbit coupling with $b^3\Pi$), $B^1\Pi$, $C^1\Sigma^+$ and $D^1\Pi$ of RbCs, and to the excitation of the core states (see text), labeled with the index {}``c". The $C_6^{Cc}(v)$ and $C_6^{Dc}(v)$ which are below 200 a.u.~for all $v$ are not plotted, for clarity. Note that the range of the $y$ axis is split into two parts with a different scale for convenience.} 
\end{center}
\end{figure*}

\newpage
\begin{figure*}
\begin{center}
\includegraphics[width=8cm]{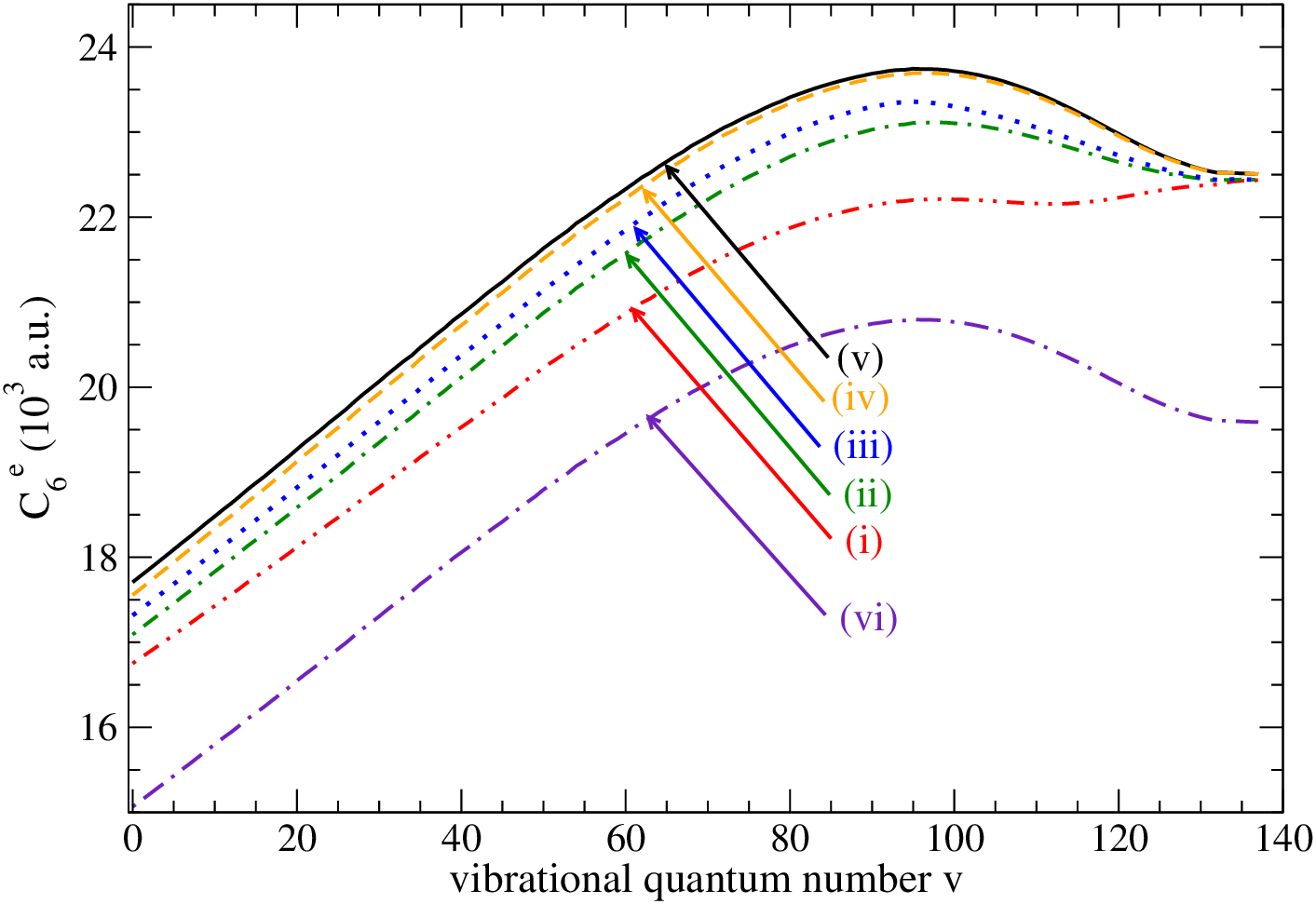}
\caption{\label{fig:C6_trans} (Color online)
Convergence of the $C_6^{e}$ coefficient between two RbCs $|X,v,j=0\rangle$ molecules with respect to the number of electronically-excited states $i'_{\textrm{max}}$ included in Eq.~(\ref{eq:c6-i1i2}), as a function of the vibrational quantum number $v$: (i) $i'_{\textrm{max}}=4$, (ii) $i'_{\textrm{max}}=6$, (iii) $i'_{\textrm{max}}=8$, (iv) $i'_{\textrm{max}}=10$, (v) $i'_{\textrm{max}}=11$. The curve (vi) corresponds to a calculation including 11 states without including the excitation of the atomic core states.} 
\end{center}
\end{figure*}

\newpage
\begin{figure*}
\begin{center}
\includegraphics[width=8cm]{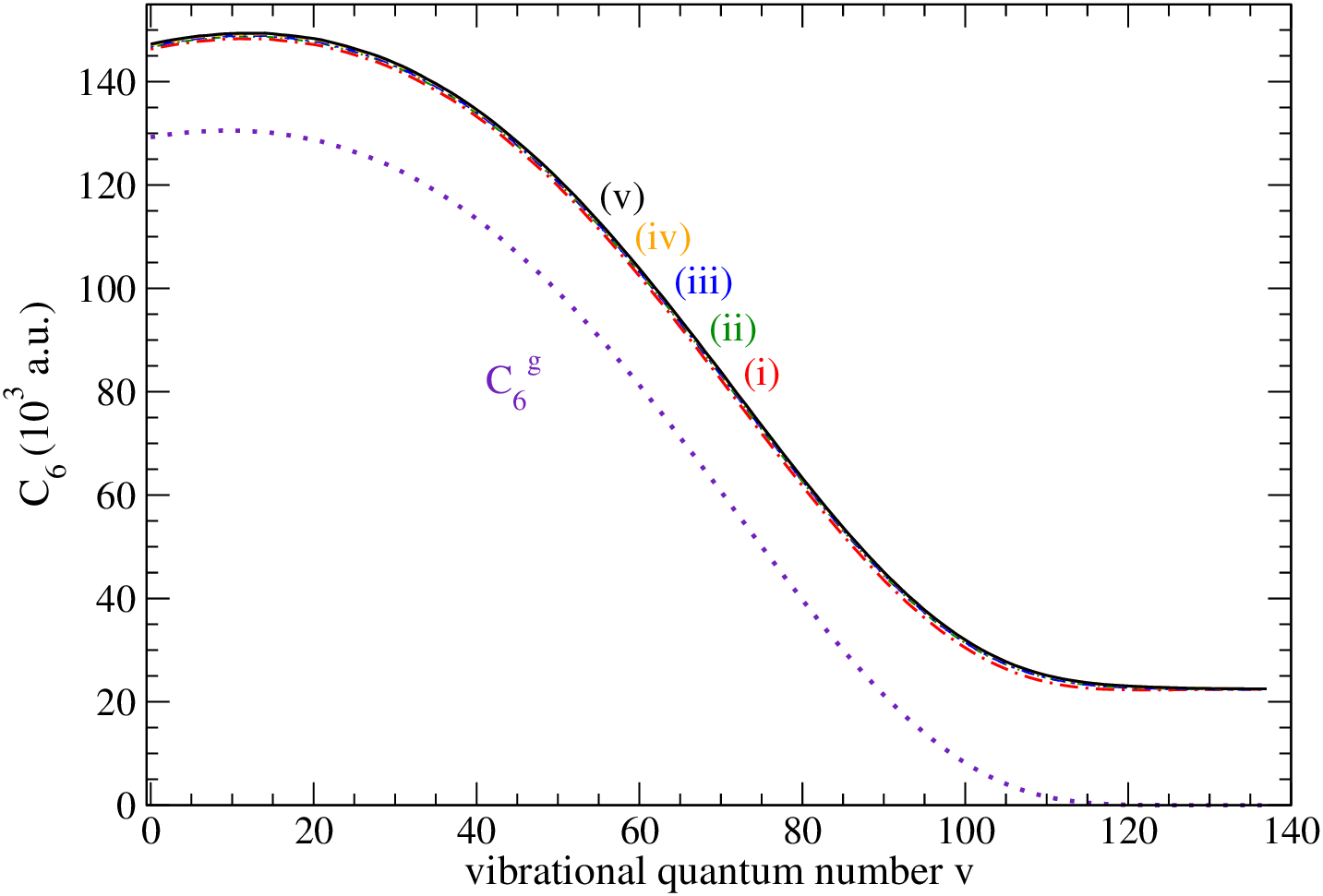}
\caption{\label{fig:C6_perm} (Color online)
Same as Figure~\ref{fig:C6_trans} for the total $C_6$ coefficient in RbCs. All curves for cases (i) to (v) are almost indistinguishable at this scale. The coefficient $C_6^\mathrm{g}(v)$ is also shown.} 
\end{center}
\end{figure*}

\newpage
\begin{figure*}
\begin{center}
\includegraphics[width=8cm]{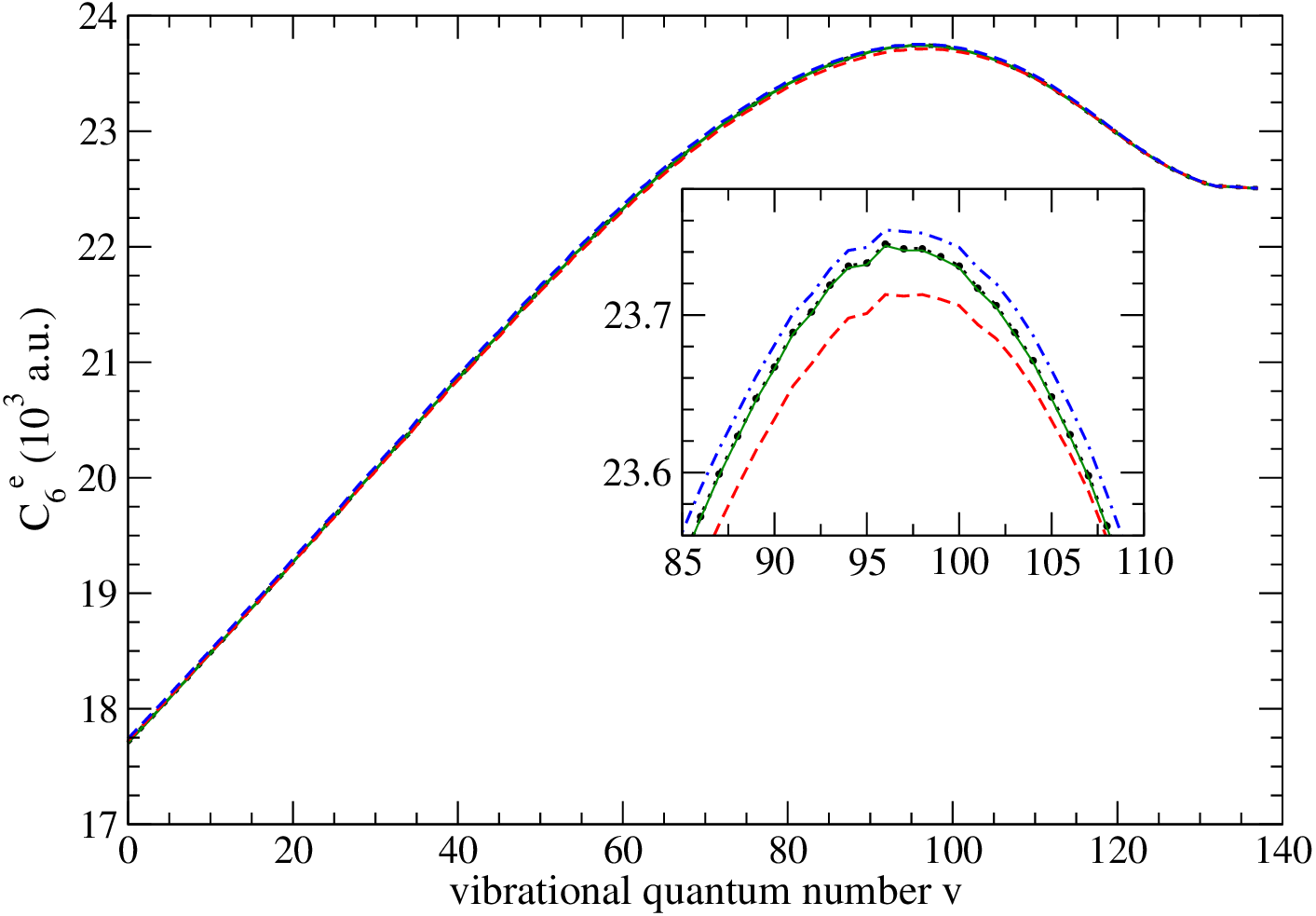}
\caption{\label{fig:C6_abinitio} (Color online) Influence on the coefficient $C_6^\mathrm{e}$ of the origin of the RbCs $A^1\Sigma^+$ potential-energy curve, as a function of the vibrational quantum number of the $X$ state. The dotted lines correspond to the RKR $A^1\Sigma^+$ potential coupled by spin-orbit interaction to the RKR $b^3\Pi$ potential; solid lines correspond to the RKR $A^1\Sigma^+$ potential without spin-orbit interaction; dashed and dash-dotted lines correspond to \textit{ab initio} potentials, respectively with and without short-range repulsive wall.} 
\end{center}
\end{figure*}

\newpage
\begin{figure*}
\begin{center}
\includegraphics[width=8cm]{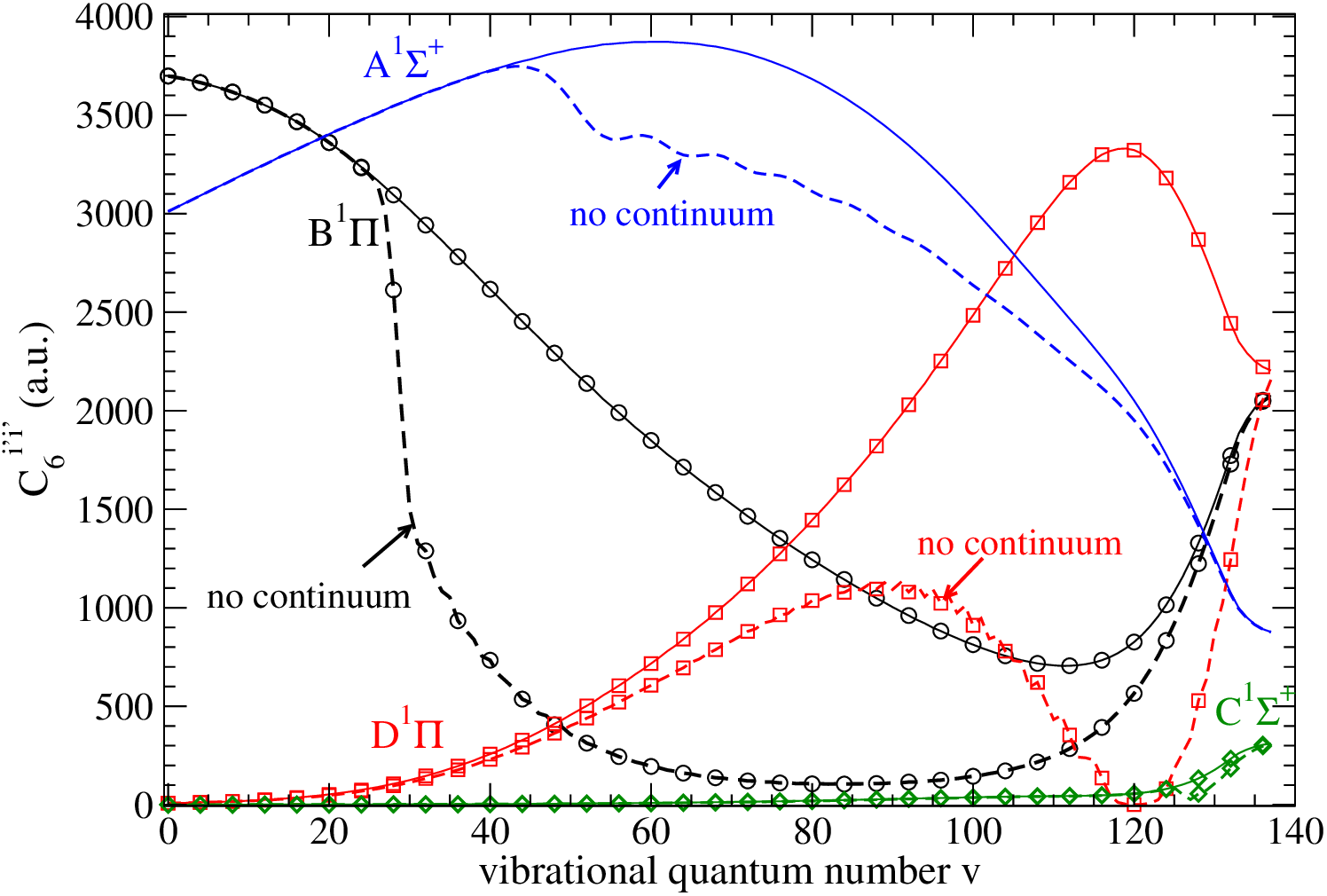}
\caption{\label{fig:C6_continuum-part} (Color online)
Influence on the partial coefficients $C_6^{i'i'}$ of the vibrational continua of the electronically-excited states $i'=A$ (without symbols), $i'=B$ (circles), $i'=C$ (squares), $i'=D$ (diamonds), as a function of the vibrational quantum number of the $X$ state. Solid lines: vibrational continua included. Dashed lines: vibrational continua not included.}
\end{center}
\end{figure*}

\newpage
\begin{figure*}
\begin{center}
\includegraphics[width=8cm]{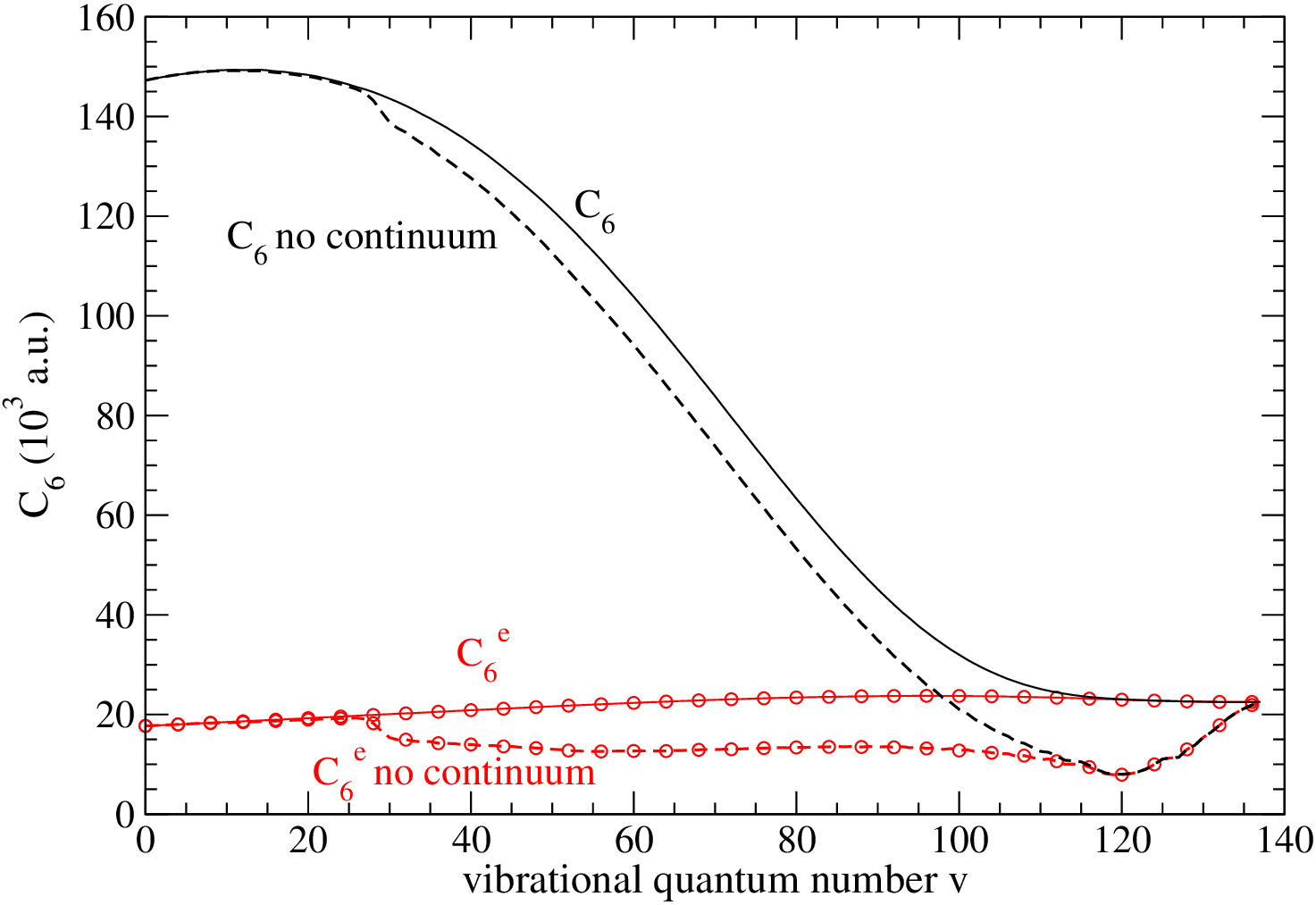}
\caption{\label{fig:C6_continuum} (Color online)
Same as Figure~\ref{fig:C6_continuum-part} for the $C_6^\mathrm{e}$ (circles) and $C_6$ (no symbols) coefficients.} 
\end{center}
\end{figure*}

\newpage
\begin{figure*}
\begin{center}
\includegraphics[width=8cm]{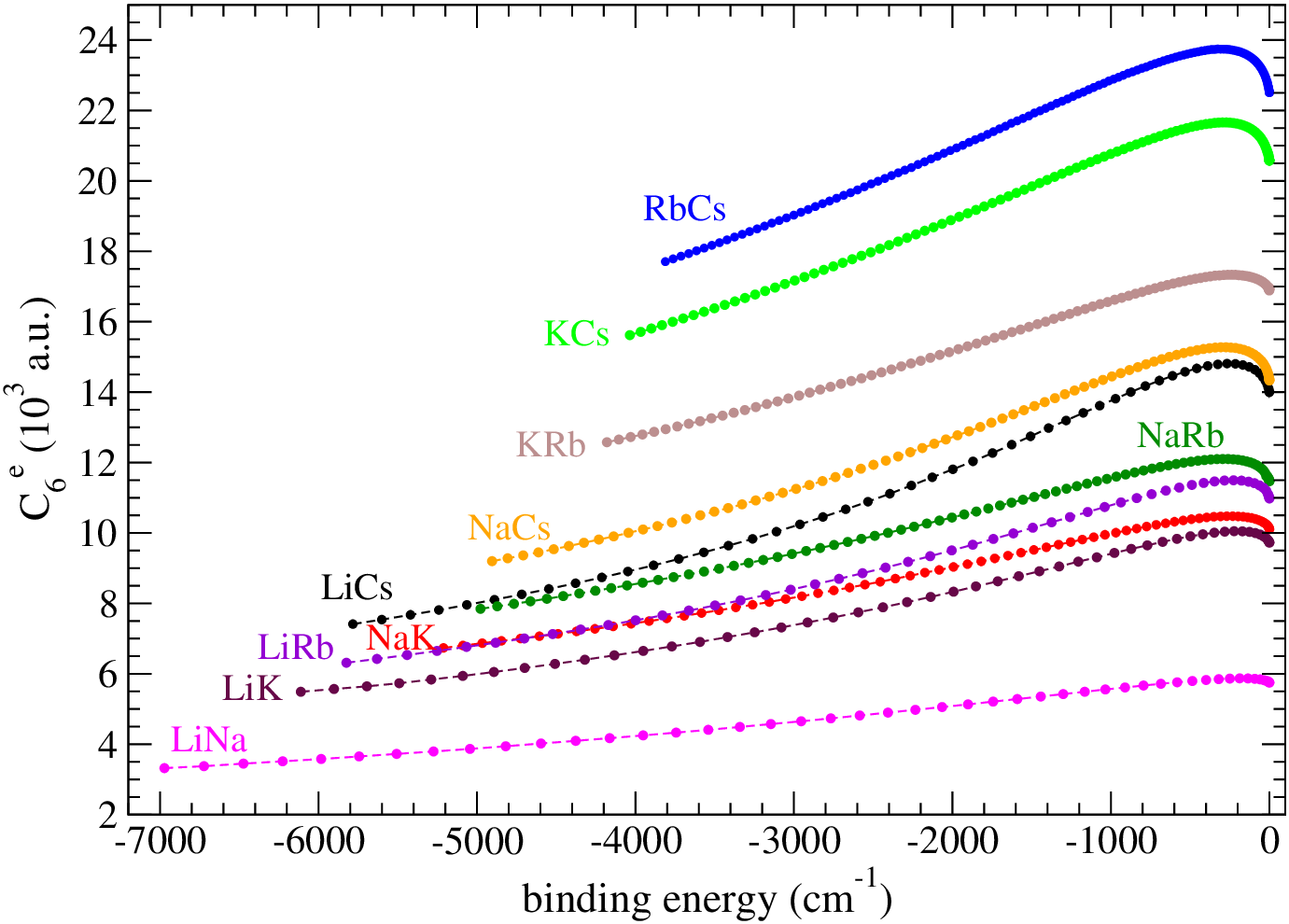}
\caption{\label{fig:C6_binding} (Color online)
Partial coefficient $C_6^\mathrm{e}$ for the interaction between two identical heteronuclear $|X^1\Sigma^+,v,j=0\rangle$ molecules as a function of the binding energy of the vibrational level $v$.}
\end{center}
\end{figure*}

\newpage
\begin{figure*}
\begin{center}
\includegraphics[width=8cm]{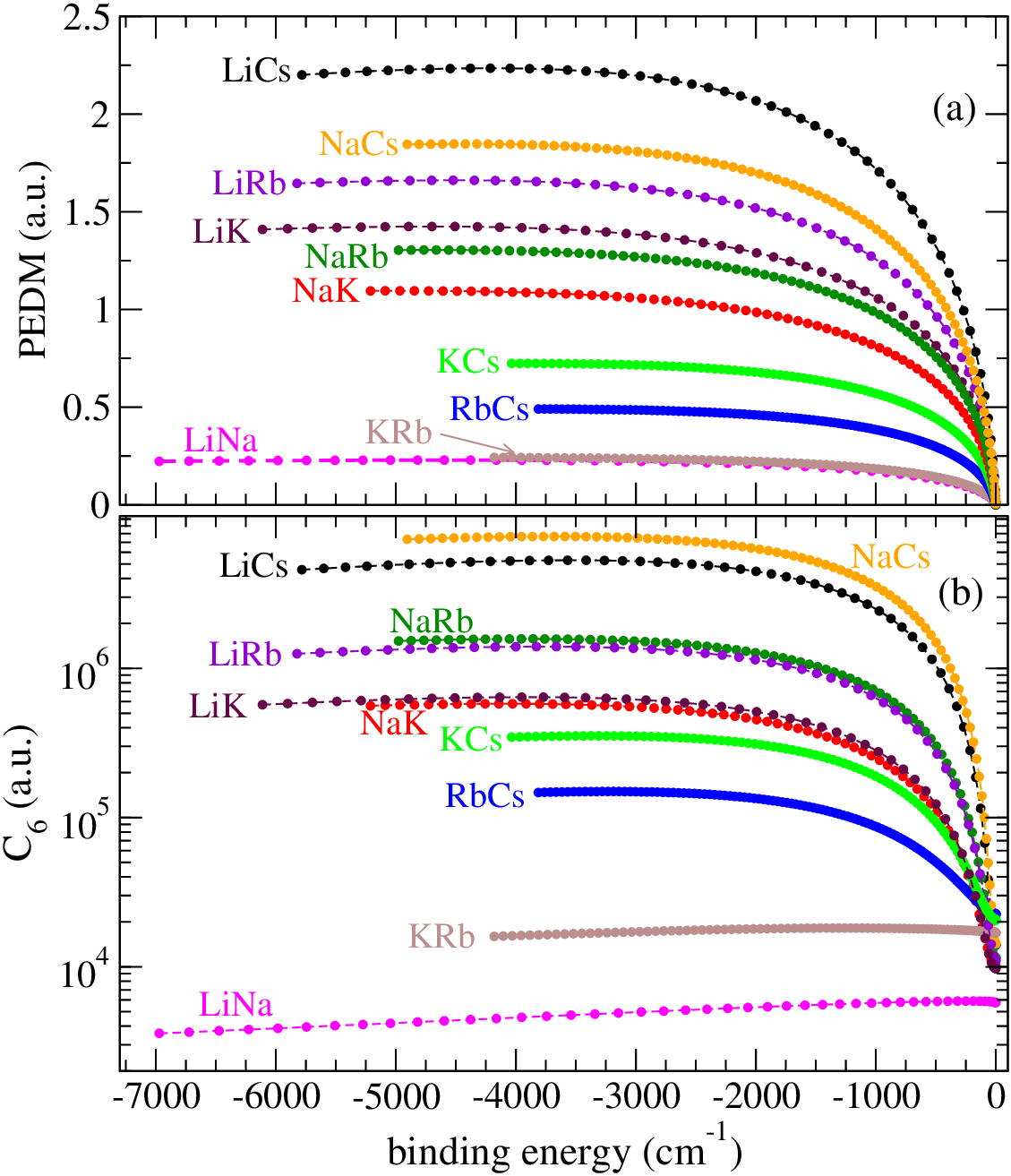}
\caption{\label{fig:C6_total} (Color online) (a) Permanent dipole moment $d_v$ of the vibrational levels of the ten species of bialkali molecules in their own frame (see Eq.~(\ref{eq:c6g-v})). (b) Total $C_6$ coefficient for the interaction between two identical heteronuclear $|X^1\Sigma^+,v,j=0\rangle$ molecules as a function of the binding energy of the vibrational level $v$.} 
\end{center}
\end{figure*}

\newpage
\begin{figure*}
\begin{center}
\includegraphics[width=8cm]{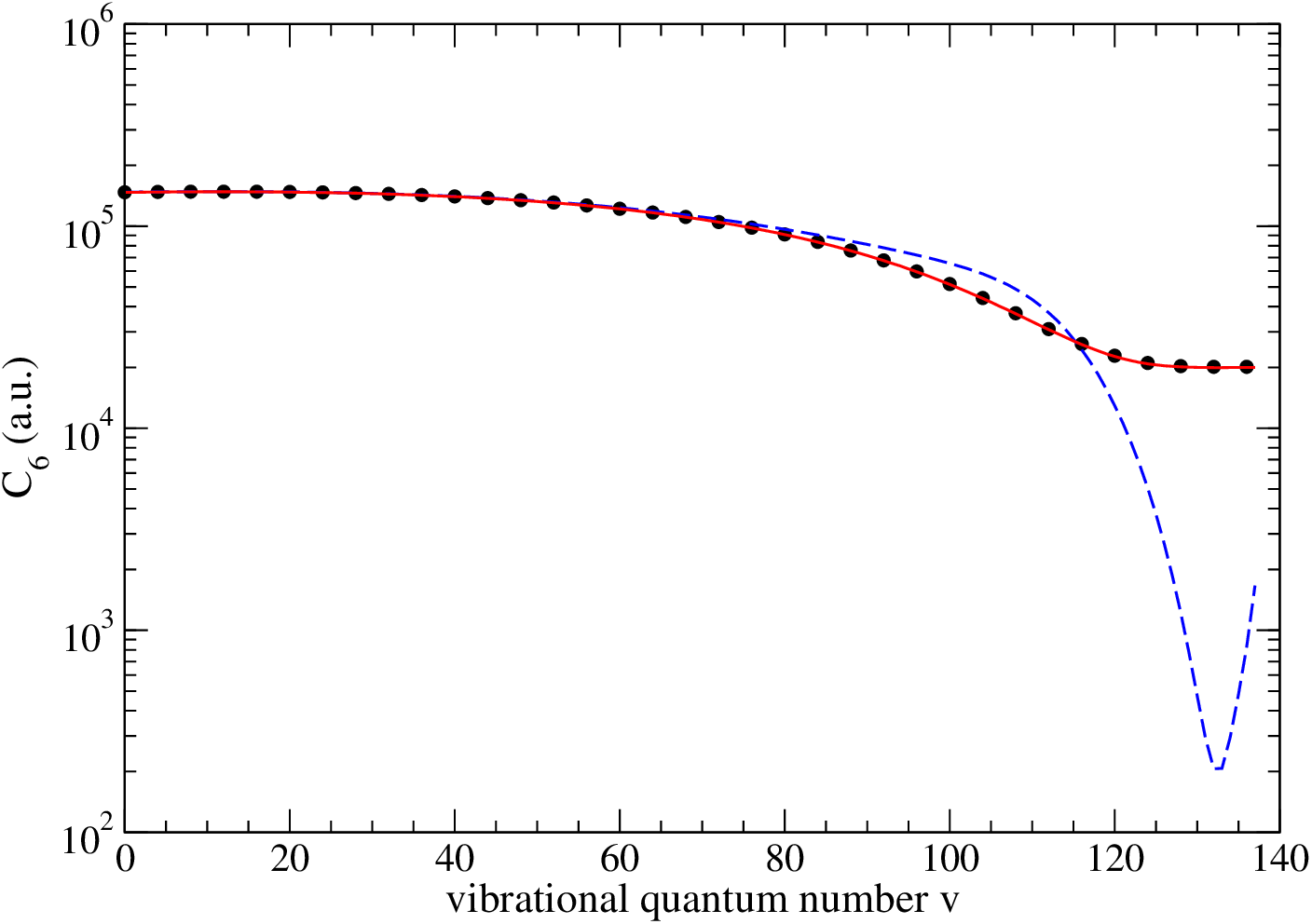}
\caption{(Color online) Total $C_6$ coefficient for the interaction between one RbCs molecule in the ground rovibronic level $|X^1\Sigma^+,v_1=0,j_1=0\rangle$, and another one in an arbitrary vibrational level $|X^1\Sigma^+,v_2=v,j_2=0\rangle$, as function of the vibrational quantum number $v$. The solid line corresponds to the numerical calculation; the dashed line corresponds to the Molwyn-Hughes approximation (\ref{eq:c6-v1v2-MH}) applied on the total $C_6$ coefficient; and the circles correspond to the Molwyn-Hughes approximation (\ref{eq:c6-v1v2-MH}) applied on each contribution $C_6^\mathrm{g}$, $C_6^\mathrm{e}$, $C_6^\mathrm{g-e}$ and $C_6^\mathrm{e-g}$ separately using Eqs.~(\ref{eq:c6-v1v2-MH-g})--(\ref{eq:c6-v1v2-MH-e-g}).
\label{fig:C6_MH}}
\end{center}
\end{figure*}

\end{document}